\pdfoutput=1

\documentclass[a4paper,12pt]{memoir}

\usepackage[hidelinks]{hyperref}
\usepackage{amsmath,amssymb,psfrag,graphicx}
\usepackage{epstopdf}
\usepackage{endnotes}

\usepackage{overpic}
\usepackage{color}
\let\footnote=\endnote

\DeclareGraphicsExtensions{eps}
\title{Maths in Medicine: How to Survive a Science Fair}
\author{Philip Pearce and Tom Shearer\\School of Mathematics, University of Manchester}
\date{}
\begin{document}

\maketitle

When talking to secondary school students, first impressions are crucial. Accidentally say something that sounds boring and you'll lose them in seconds. A physical demonstration can be an eye-catching way to begin an activity or spark off a conversation about mathematics. This is especially true in the context of an event like a science fair where there are hundreds of other exhibitors and stands, possibly involving loud music and/or dancing robots!

In this article we describe three devices that were built to illustrate specific physical phenomena that occur in the human body. Each device corresponds to a simple mathematical model which contains both elements that are accessible to pupils in the early years of secondary education and more challenging mathematical concepts that might appeal to A-level students. Two of the devices relate to the Windkessel effect, a physical phenomenon that regulates blood flow, and the third demonstrates the elastic properties of ligaments and tendons.

\begin{figure}
\begin{centering}
\begin{overpic}[width=\textwidth]{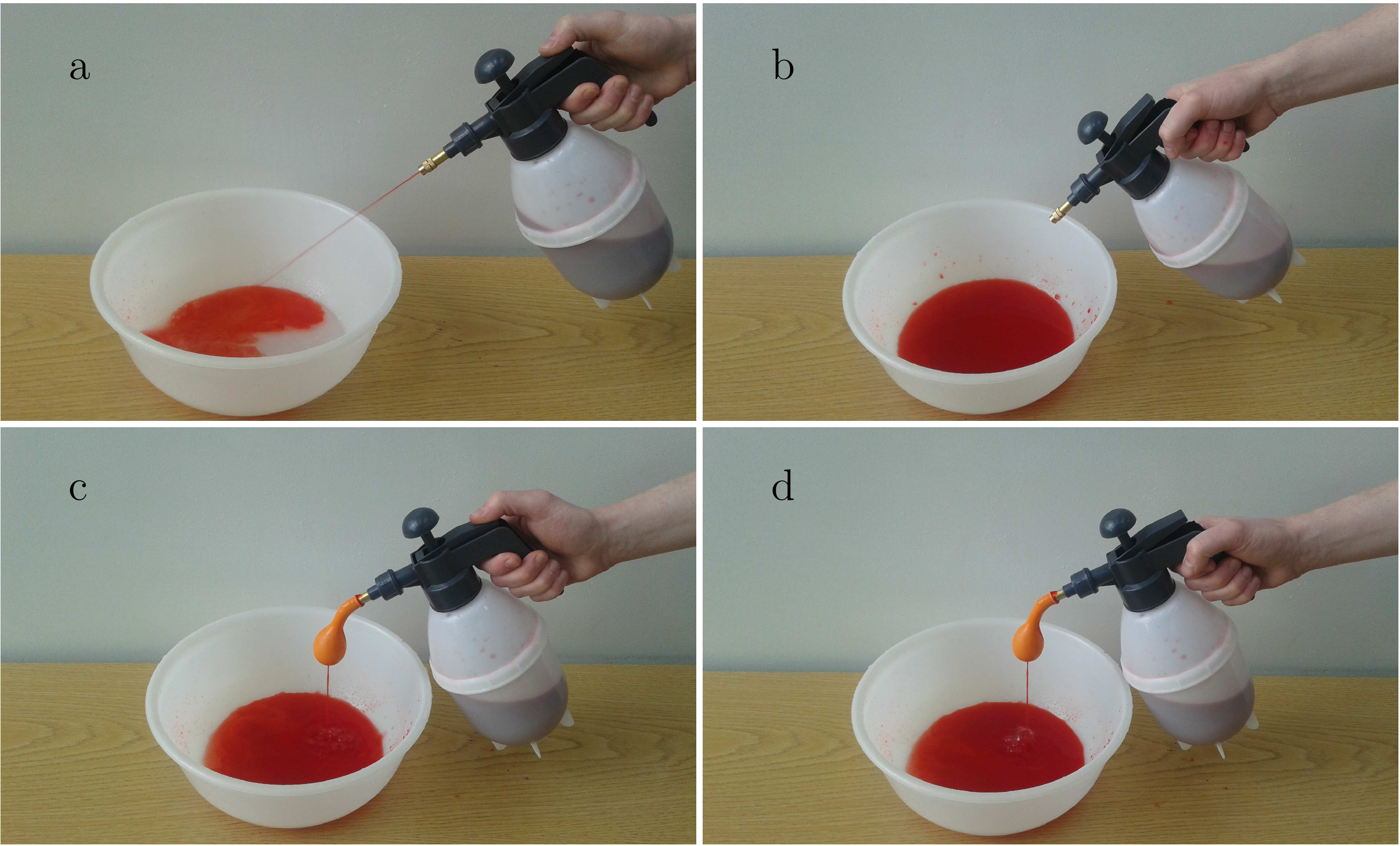}
\end{overpic}
\caption{A demonstration of the Windkessel effect using a garden sprayer and a balloon (red food colouring optional).}\label{fig:sprayer}
\end{centering}
\end{figure}

\begin{figure}
\begin{centering}
\begin{overpic}[width=\textwidth]{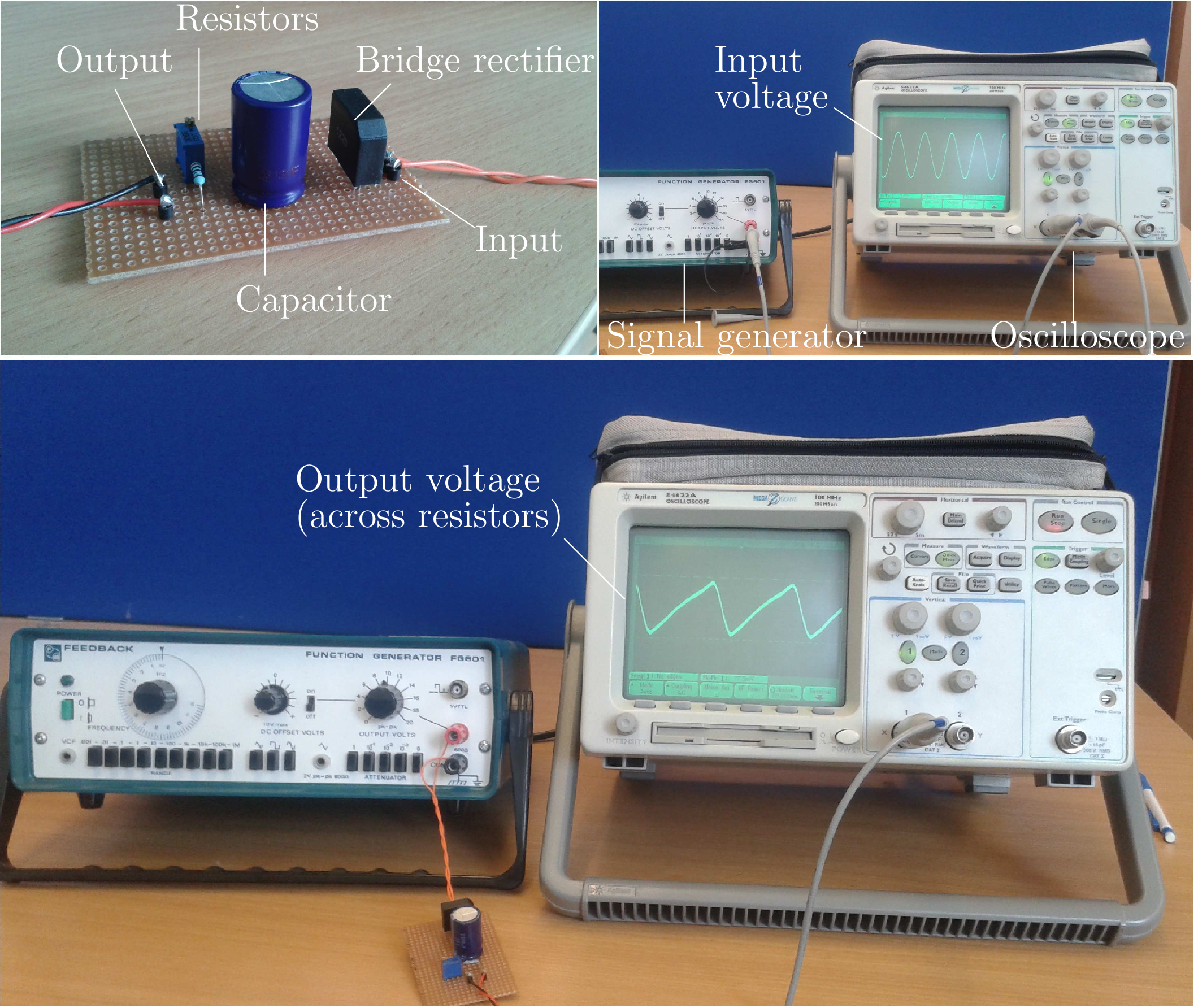}
\end{overpic}
\caption{A demonstration of a capacitor smoothing potential difference across two resistors (one is a variable resistor for flexibility).\setcounter{endnote}{0}\protect\endnotemark ~The bridge rectifier ensures that only positive voltages (i.e. the peaks of the sinusoidal input) are transmitted to the circuit \cite{circuit}; therefore the transmitted voltage is similar to the input current plotted in Figure \ref{fig:flowcomparison}.}\label{fig:oscillator}
\end{centering}
\end{figure}
\endnotetext{The circuit shown in Figure \ref{fig:oscillator} is slightly more complicated than the one modelled in the following section. In Figure \ref{fig:oscillator} the input voltage is imposed with a signal generator and a bridge rectifier, due to the difficulty of imposing the current; however, the mechanism, whereby a capacitor smooths potential difference across a resistor, is similar. An interactive simulation of the circuit shown in Figure \ref{fig:oscillator} can be found in \cite{simulation}.}
\section*{Hearts and arteries}
\label{sec:windkessel}
When the heart beats, it generates sharp spikes in blood pressure. If the arteries connected to the heart were rigid, these pressure spikes would be transmitted to the circulatory system, potentially causing serious physiological problems \cite{aorta}. In addition, because the flow of blood out of the heart stops periodically (during diastole), the heart would have to work hard to accelerate the entire mass of blood within the circulatory system each time it beats. These effects are prevented thanks to the compliance of the aorta - the blood vessel that connects the heart to the rest of the circulatory system. In the presence of a fluctuating flow rate, it stores and slowly releases blood, thus damping the oscillations in pressure and allowing a continuous blood flow downstream. This phenomenon is often called the Windkessel effect \cite{aorta}. In this section we describe two devices that analogously illustrate this process, and two corresponding mathematical models.

\subsection*{Balloons and circuits}

The Windkessel effect can be simply demonstrated using a garden sprayer and a small balloon, as shown in Figure \ref{fig:sprayer}. Squeezing the handle of the unmodified sprayer at regular intervals produces sharp bursts of fluid flow (Figures \ref{fig:sprayer}a and \ref{fig:sprayer}b), whereas when a balloon is attached to its nozzle, the same input produces a continuous flow (Figures \ref{fig:sprayer}c and \ref{fig:sprayer}d). For an excellent video demonstration of a similar device, see \cite{windkesselvid}.

A similar phenomenon occurs in a completely different area of physics (see Figure \ref{fig:oscillator}). The figure shows a circuit with a capacitor in parallel with a resistor; the capacitor damps oscillations in potential difference through the resistor in a process that is analogous to the way that a compliant tube damps fluid flow pressure fluctuations. The mathematical equations governing electric circuits like the one shown in Figure \ref{fig:oscillator} have been used to study the heart and circulation, as discussed below.

\subsection*{Mathematical models}

The devices described above can be modelled as follows. Consider an incompressible, viscous fluid flowing steadily through a rigid pipe of radius H. Assuming the flow is axisymmetric, fully developed and flowing only in the direction parallel to the pipe, the Navier-Stokes equations, which govern viscous fluid flow, reduce to 
\begin{equation}
\frac{1}{r}\frac{\partial}{\partial r}\left(r \frac{\partial u_z}{\partial r}\right) = \frac{1}{\mu}\frac{d p(z)}{d z}. \label{ns}
\end{equation}
Here $\mu$ is the dynamic viscosity of the fluid, $u_z$ is its velocity along the pipe, $r$ and $z$ are radial and longitudinal variables, respectively, and $p$ is the fluid pressure. Integrating \eqref{ns} twice with respect to $r$, applying a no-slip condition $(u_z=0)$ on the boundary $r=H$ and imposing the condition that $u_z$ remains finite at $r=0$, gives
\begin{equation}
u_z = -\frac{1}{4\mu}\frac{d p(z)}{d z}\left(H^2-r^2\right), \label{rmomentum}
\end{equation}
which is known as Hagen--Poiseuille flow. Integrating \eqref{rmomentum} across the pipe cross-section gives the volumetric flow rate
\begin{equation}
Q = -\frac{\pi H^4}{8\mu}\frac{d p(z)}{d z}. \label{flowrate}
\end{equation}
Finally, integrating \eqref{flowrate} along the length $L$ of the pipe leads to Poiseuille's law

\begin{equation}
P = Q\left(\frac{8\mu L}{\pi H^4}\right) = QR, \label{poiseuille}
\end{equation}
where $P$ is the pressure drop along the pipe.

Similarly, Ohm's law relates the voltage (or potential difference) $V$ across a resistor to its current $I$ and resistance $R$:
\begin{equation}
V = IR.
\end{equation}
As many secondary school pupils know, explanations of Ohm's law often use the analogy of water in pipes \cite{ohms}. What many of them aren't aware of, however, is that, over the past hundred years or so, researchers have used the flow of electric charge in circuits as an analogy for studying the circulation of blood around the body.

Modelling capillaries in the circulatory system as a network of resistors in an electric circuit is a simple and useful way of studying how blood flows through the human body. However, the assumption of zero flow velocity at the capillary wall, used in the derivation of Poiseuille's law, is only valid in blood vessels with walls that behave as if they are rigid\footnote{In practice, due to the assumption of fully developed flow, Poiseuille's law only applies in thin blood vessels, where the Reynolds number of the flow is small.}.

The large arteries connecting the left ventricle of the heart to the rest of the circulatory system are particularly compliant, so Poiseuille's law does not hold. They fill up with blood as the heart beats (systole) and slowly release blood as the heart refills (diastole), as described above. If we are to continue with the electrical analogy, we require an additional component in the circuit - a capacitor.

\begin{figure}
\begin{overpic}[width=\textwidth]{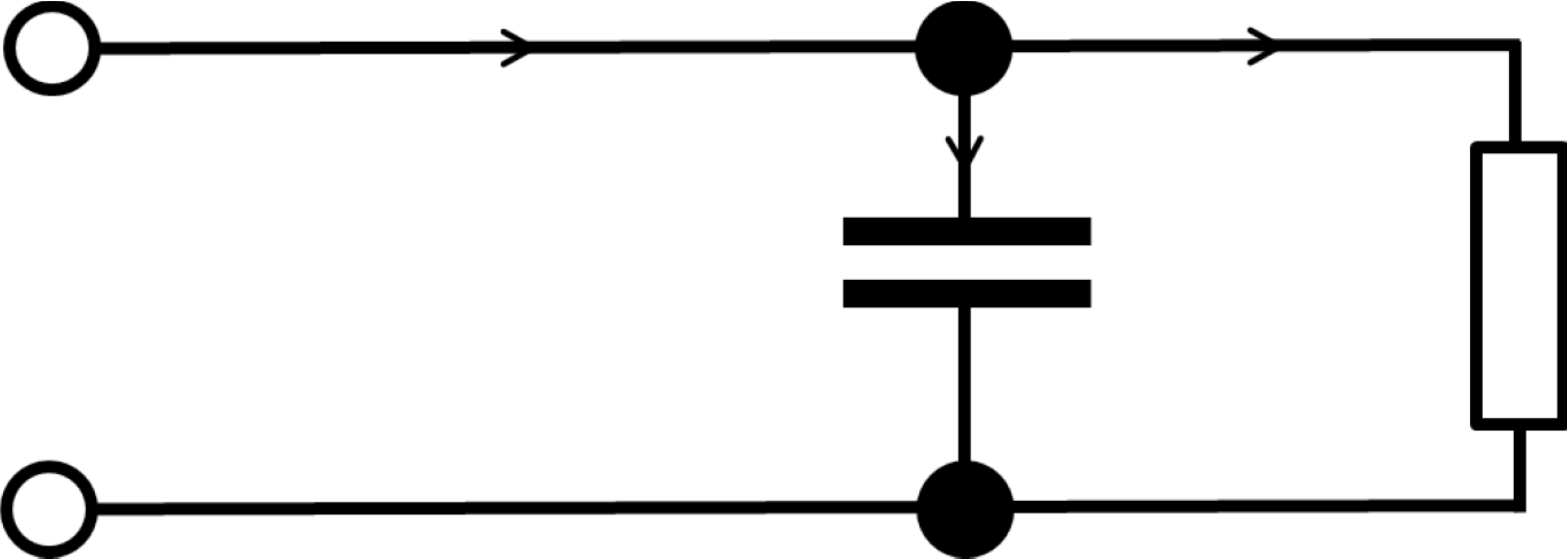}
\put (30,27) {\Large $I$}
\put (56,25) {\Large $I_1$}
\put (80,27) {\Large $I_2$}
\end{overpic}
\caption{A circuit diagram of a capacitor in parallel with a resistor.}\label{fig:circuit}
\end{figure}

A capacitor in parallel with a resistor, with a fluctuating current applied to the circuit, stores and releases voltage to the resistor \cite{circuit} in a manner that is similar to the way in which elastic arteries store and release blood to the rest of the circulatory system.  The equation relating voltage to current in the circuit shown in Figure \ref{fig:circuit} can be found by noting that the total current is given by
\begin{equation}
I = I_1 + I_2, \label{current}
\end{equation}
where
\begin{equation}
I_1=C\frac{d V}{d t}\quad\text{and}\quad I_2=\frac{V}{R}.
\label{current2}
\end{equation}
The first of these is the equation for a capacitor (where $C$ is its capacitance and $t$ is time); the second is Ohm's law. Inserting \eqref{current2} into \eqref{current} gives
\begin{equation}
C \frac{d V}{d t} + \frac{V}{R} = I.
\end{equation}
The equivalent equation for blood flow is
\begin{equation}
C \frac{d P}{d t} + \frac{P}{R} = Q, \label{pressureode}
\end{equation}
where, in this case, $C$ is the compliance of the aorta and $R$ is the total resistance of the rest of the circulatory system\footnote{Here we have made the simplifying assumptions that the aorta is the only compliant artery and that the rest of the circulation obeys Poiseuille's law.}. Assuming we can measure the flow rate $Q$ out of the heart, we can find the pressure drop $P$ from equation \eqref{pressureode} using the integrating factor method to give
\begin{figure}
\begin{overpic}[width=\textwidth]{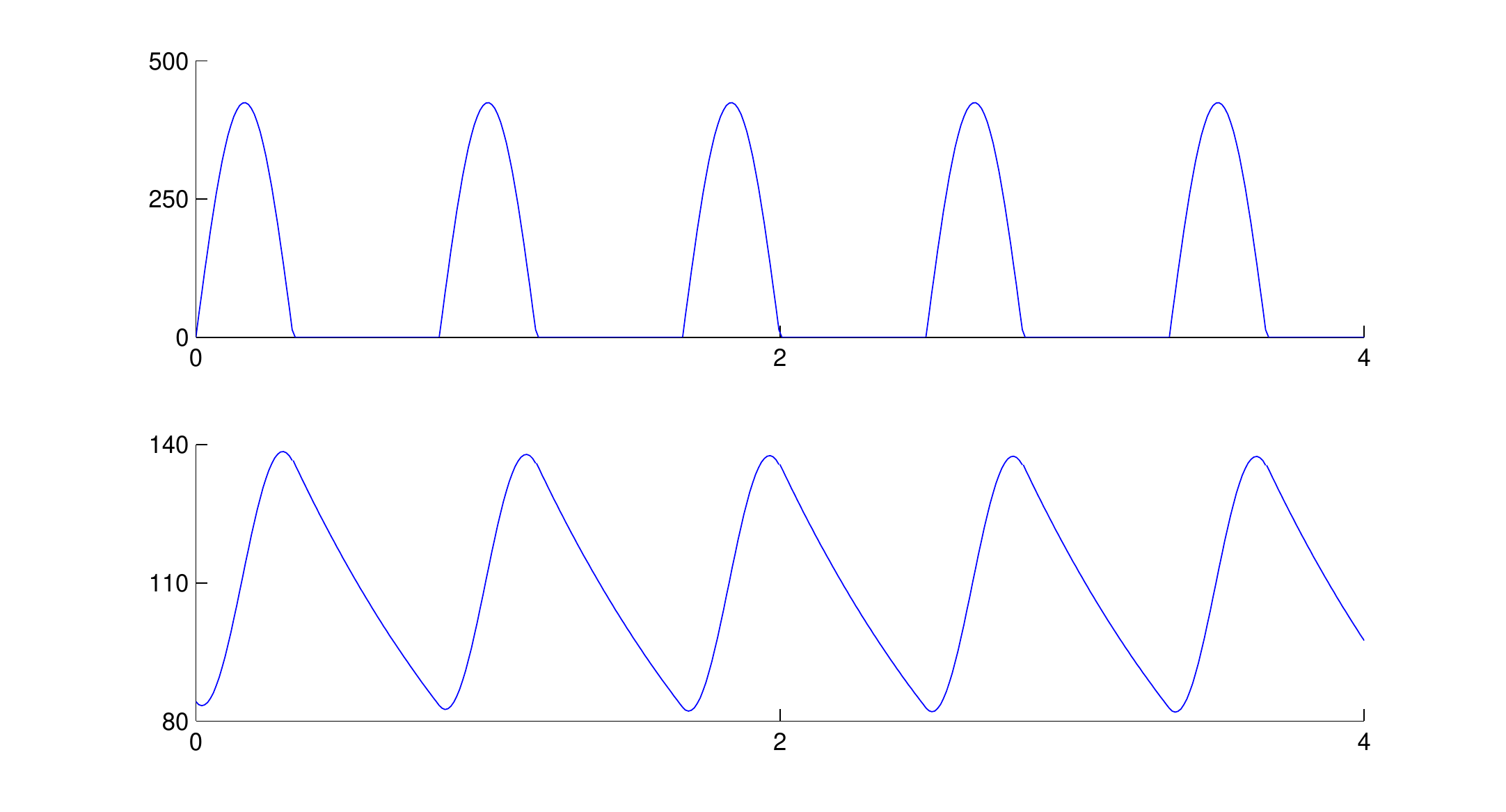}
\put (47,0) {time (s)}
\put(0,27){\small$Q_2$ (ml/s)}
\put(0,53){\small $Q$ (ml/s)}
\end{overpic}
\caption{Comparison of flow out of the heart $Q$ with flow through the circulatory system $Q_2$. The code used to produce this figure is similar to that used in \cite{windkessel2}.}\label{fig:flowcomparison}
\end{figure}
\begin{equation}
P(t) = \exp\left({-\frac{t}{RC}}\right)\int \frac{\exp\left({\frac{t}{RC}}\right) Q(t)}{C}~\mathrm{d}t. \label{pressure}
\end{equation}
We can then use this equation to plot the flow through the circulatory system, which, using Poiseuille's law \eqref{poiseuille}, is given by
\begin{equation}
Q_2 = \frac{P(t)}{R}. \label{q2}
\end{equation}
For example, assume the flow out of the heart is given by
\begin{equation}
Q=Q_0 \sin \left(\frac{\pi t}{t_0}\right) \label{flowheart}
\end{equation}
during systole, which is of length $t_0$, where $Q_0$ is the maximum amplitude of the flow rate. Then during systole we have, using \eqref{pressure} and \eqref{flowheart}
\begin{equation}
P_s(t) =  \frac{t_0Q_0R\left(t_0 \sin\left(\frac{\pi t}{t_0}\right)-RC\pi\cos\left(\frac{\pi t}{t_0}\right)\right)}{\pi^2C^2R^2+t_0^2}+K_1\exp\left(-\frac{t}{RC}\right), \label{systole}
\end{equation}
where $K_1$ is a constant of integration. Assuming there is no flow during diastole, we have
\begin{equation}
P_d(t) = K_2 \exp \left(-\frac{t}{RC}\right), \label{diastole}
\end{equation}
where $K_2$ is another constant of integration. The constants $K_1$ and $K_2$ can be found by enforcing continuity between systolic and diastolic pressures. Using physiological values for $Q_0$, $t_0$, $R$ and $C$ taken from \cite{windkessel2}, we can plot the flow out of the heart $Q$, given by \eqref{flowheart}, and the flow through the circulatory system $Q_2$ (see Figure \ref{fig:flowcomparison}), found by inserting \eqref{systole} and \eqref{diastole} into \eqref{q2}. The figure clearly shows the continuous flow through the circulatory system, despite the discontinuous flow out of the heart.
\begin{figure}
\begin{center}
\begin{overpic}[scale=1]{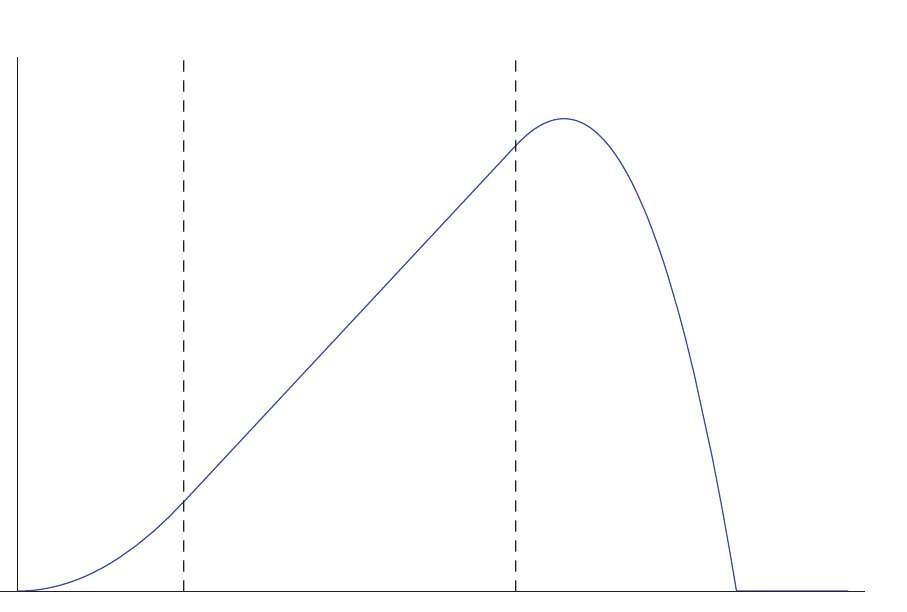}
\put(-4,62){stress}
\put(90,-2){strain}
\put(8,48){\textit{toe}}
\put(6,44){\textit{region}}
\put(25,48){\textit{linear region}}
\put(76,48){\textit{failure region}}
\end{overpic}
\caption{Schematic representation of typical ligament or tendon stress--strain behaviour.}
\label{typical}
\end{center}
\end{figure}
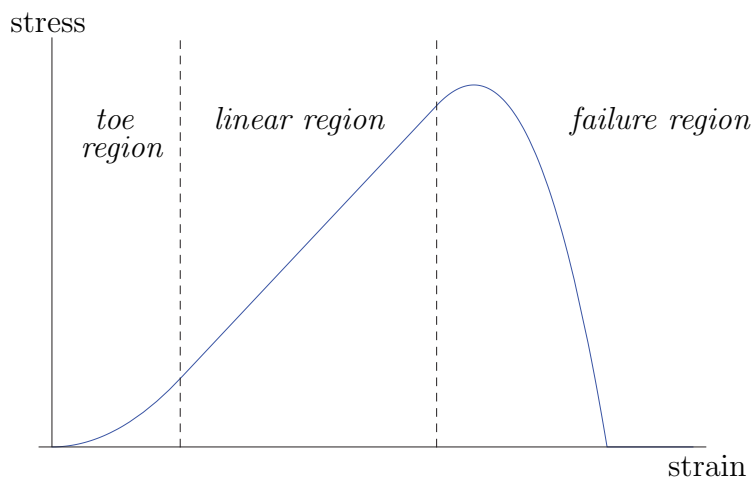
\section*{Ligaments and tendons}
\label{sec:tendons}
Ligaments and tendons are important connective tissues in the human body. Ligaments connect bone to bone to provide stability, whereas tendons connect bone to muscle in order to facilitate the transfer of forces from the muscles to the skeleton. They exhibit complex mechanical behaviour, which is the subject of a large body of current research \cite{RW15}.

\begin{figure}
\begin{center}
\begin{overpic}[scale=0.6]{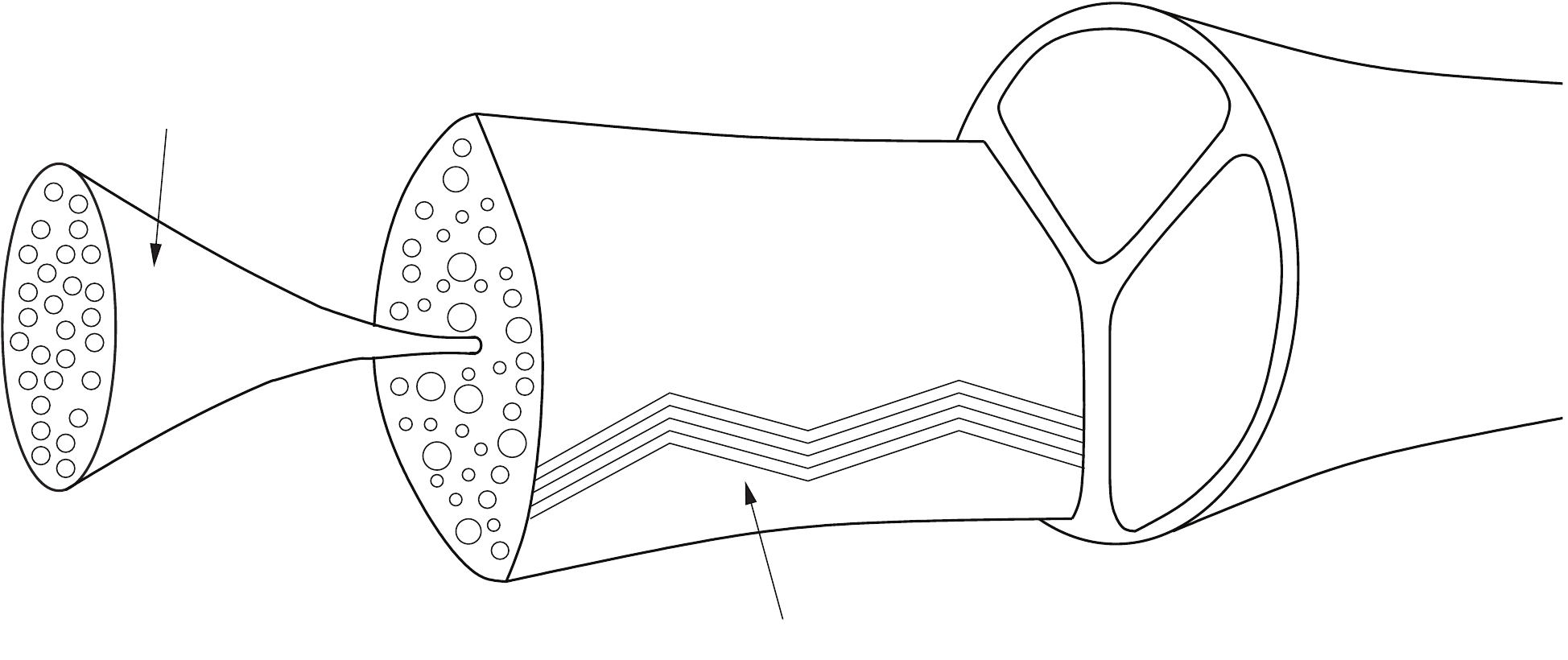}
\put(7,35){\textit{fibril}}
\put(46,-0.5){\textit{crimp}}
\put(43,22){\textit{fascicle}}
\put(86,25){\textit{tendon}}
\end{overpic}
\caption{Idealised tendon hierarchy (adapted from \cite{S15}).}
\label{tendon_hierarchy_fig}
\end{center}
\end{figure}

When subjected to uniaxial extension, ligaments and tendons exhibit a distinctive stress--strain response (see Figure \ref{typical}), which is initially non-linear with increasing stiffness (this region of the curve is termed the \textit{toe region}), and subsequently linear after some critical strain. This can be explained by considering their internal structure. Ligaments and tendons have a complex hierarchy of subunits (see Figure \ref{tendon_hierarchy_fig}). They consist of long, thin fibres called fascicles, which in turn are made of thinner fibres called fibrils, which are initially crimped and have differing lengths. As a tendon is stretched, the fibril crimp begins to straighten out and individual fibrils start to tauten. As more and more fibrils become taut, the tendon gets stiffer and stiffer until all the fibrils are taut, from which point its stiffness remains constant.

\begin{figure}
\begin{centering}
\includegraphics{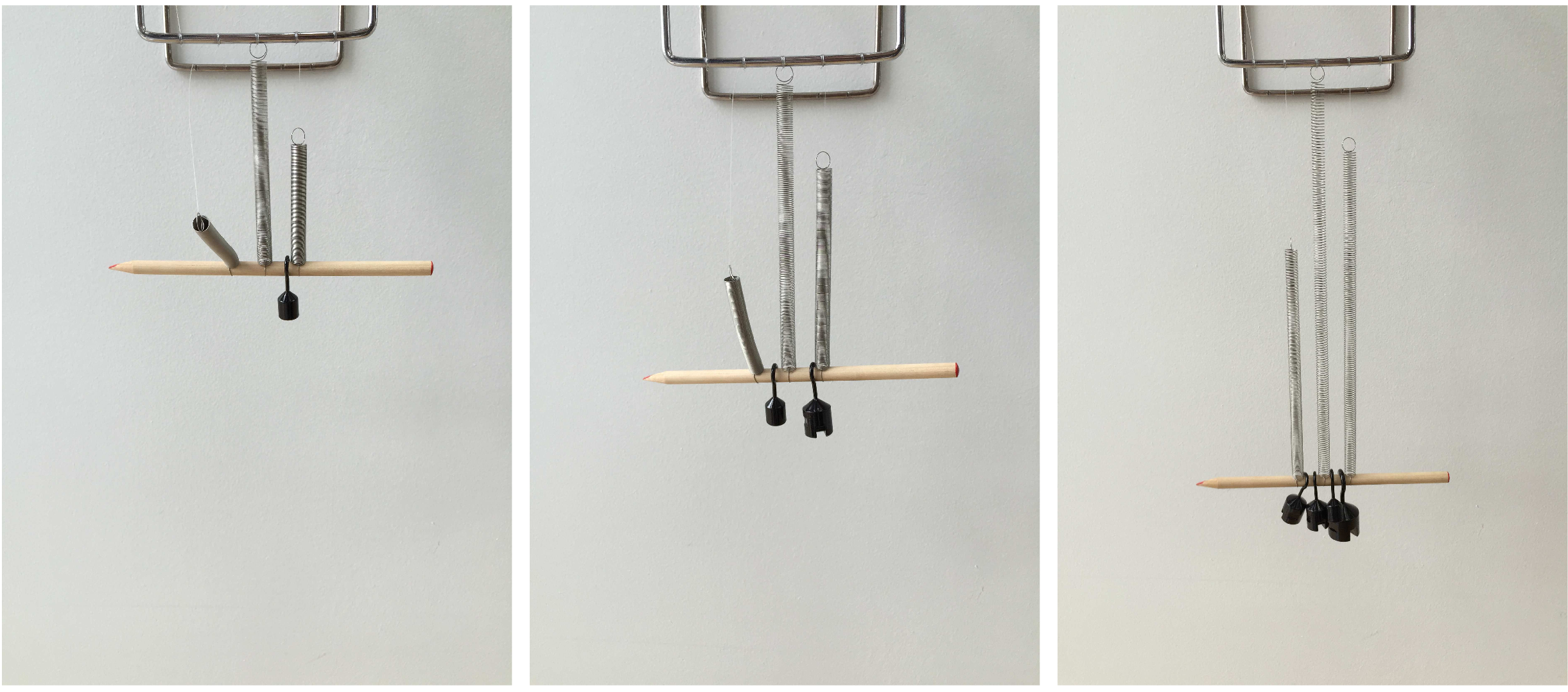}
\caption{A demonstration of the behaviour of ligaments and tendons using springs and strings.}
\label{fig:springs}
\end{centering}
\end{figure}
\subsection*{Springs and strings}

The behaviour described above can be illustrated via a system of springs and strings (see Figure \ref{fig:springs}). The springs (which all have the same undeformed length) represent the elasticity of the fibrils, whereas the strings (which have differing lengths) represent their crimp. The strings ensure that the springs they are attached to do not contribute to the system's stiffness as a whole until they are fully straightened out, thus simulating what occurs in a ligament or tendon. 

\subsection*{Mathematical model}

Whilst the mathematics of the fully non-linear, time-dependent mechanical behaviour of a tendon would be overwhelming to a secondary school student, a simple model can be used to explain the concept in a way that is easily accessible.

Consider a tendon subject to a uniaxial strain ($e$), defined as:
\begin{equation}
e=\frac{\Delta L}{L},
\end{equation}
where $\Delta L$ is the change in length, and $L$ is its initial length. Stress ($\sigma$) is defined as the force ($F$) per unit area ($A$) required to impose the strain:
\begin{equation}
\sigma=\frac{F}{A}.
\end{equation}

The behaviour described above can be explained mathematically via the sequential straightening and loading model \cite{KPB80}. In this model, each fibril is assumed to obey Hooke's law:
\begin{equation}
\sigma_f=Ee_f,
\label{HookesLaw}
\end{equation}
where $\sigma_f$ is the fibril stress, $e_f$ is the fibril strain and $E$ is the fibril Young's modulus, and by considering a fibril of given initial length $l$, the strain in that fibril can be calculated as a function of the applied tendon strain.

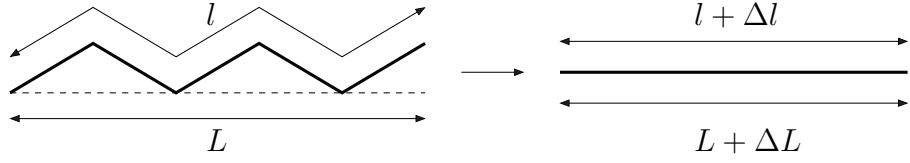
\begin{figure}
\begin{center}
\begin{overpic}[scale=0.43]{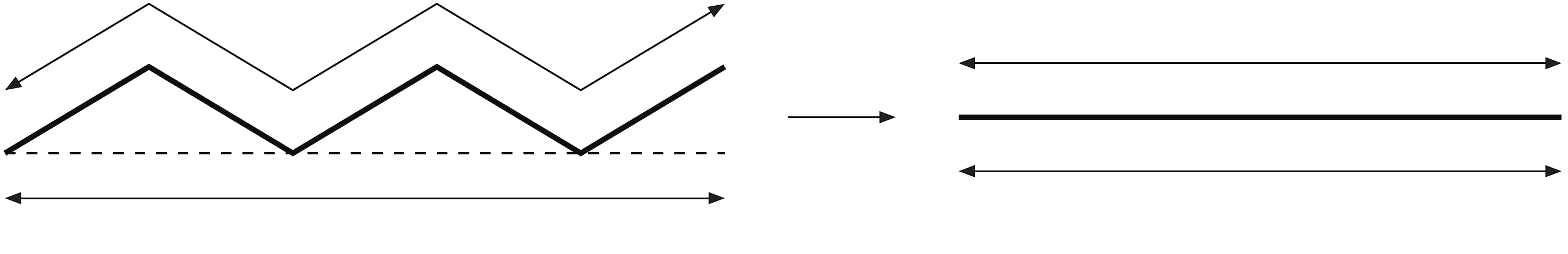}
\put(22,14){$l$}
\put(22,0){$L$}
\put(76,14){$l+\Delta l$}
\put(76,0){$L+\Delta L$}
\end{overpic}
\caption{A fibril of initial length $l$ within a tendon of initial length $L$ is stretched beyond the point at which it becomes taut, so that $l+\Delta l=L+\Delta L$.}
\label{fibstrain}
\end{center}
\end{figure}

It is assumed that a fibril does not experience any strain until its crimp has fully straightened out, from which point the current fibril and tendon lengths will be the same (see Figure \ref{fibstrain}):
\begin{align}
l+\Delta l&=L+\Delta L \\
\Rightarrow\frac{l+\Delta l}{L}&=1+e \\
\Rightarrow\frac{l}{L}(1+e_f)&=1+e \\
\Rightarrow e_f&=\frac{L}{l}(1+e)-1.
\end{align}

Therefore,
\begin{equation}
e_f=\begin{cases}
0, & e<\frac{l}{L}-1,\\
\frac{L}{l}(1+e)-1, & e\ge\frac{l}{L}-1. 
\end{cases}
\label{FibrilStrain}
\end{equation}
Hence, assuming the ratio of the initial fibril length to the initial tendon length is known, the fibril strain (and hence stress) can be calculated as a function of the tendon strain. Since $l$ is different for each fibril, the stress in each fibril will also be different. The total tendon stress can then be calculated as the average of the stresses in all of its fibrils. 

\begin{figure}
\begin{center}

\begin{overpic}[scale=1]{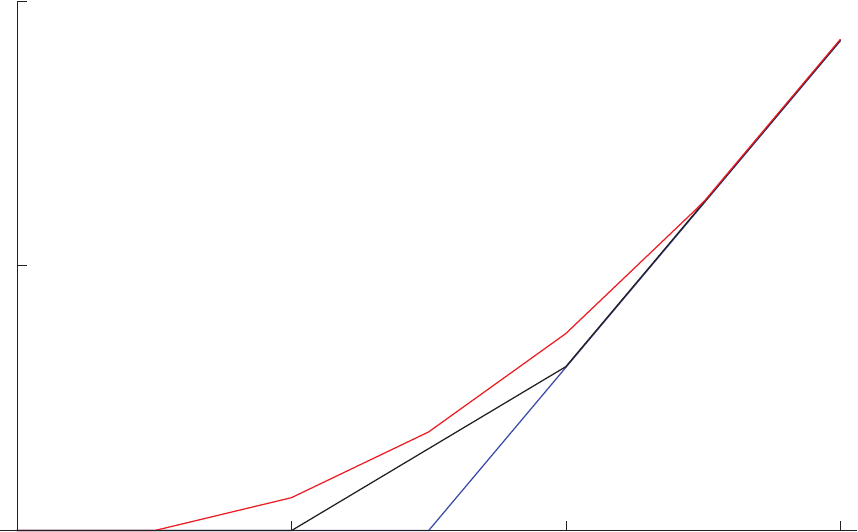}
\put(-5,65){$\sigma$ (MPa)}
\put(-4,59){\small{80}}
\put(-4,29.5){\small{40}}
\put(1,-4){\small{0}}
\put(33.3,-4){\small{1}}
\put(65,-4){\small{2}}
\put(96.5,-4){\small{3}}
\put(101,-2){$e$ (\%)}
\end{overpic}
\caption{Fascicle stress as a function of strain for a fascicle with one (blue), two (black) and five (red) fibrils.}
\label{stress}
\end{center}
\end{figure}
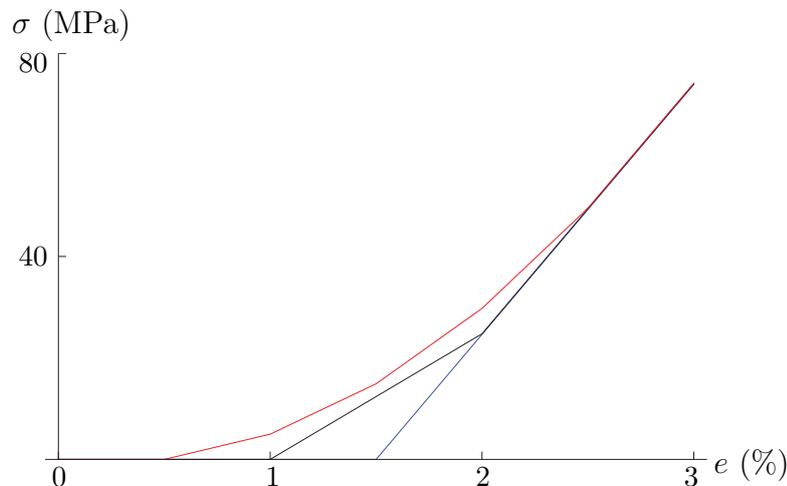

As an example, let us consider an idealised tendon that consists of only two fibrils, and assume that the first fibril has initial length $l_1=101L/100$ and the second has initial length $l_2=102L/100$. Using equations \eqref{HookesLaw} and \eqref{FibrilStrain}, the stresses in the first and second fibrils, respectively, are:
\begin{equation}
 \sigma_f^{(1)}=\begin{cases}
0, & e<\frac{1}{100},\\
E\left(\frac{100}{101}(1+e)-1\right), & e\ge\frac{1}{100},
\end{cases}
\end{equation}
\begin{equation}
\sigma_f^{(2)}=\begin{cases}
0, & e<\frac{2}{100},\\
E\left(\frac{100}{102}(1+e)-1\right), & e\ge\frac{2}{100}.
\end{cases}
\end{equation}
The total stress experienced by the tendon, therefore, is
\begin{equation}
 \sigma=\frac{\sigma_f^{(1)}+\sigma_f^{(2)}}{2}=\begin{cases}
0, & e<\frac{1}{100},\\
E\left(\frac{50}{101}(1+e)-\frac{1}{2}\right), & \frac{1}{100}<e\le\frac{2}{100},\\
E\left(\frac{5075}{5151}(1+e)-1\right), & e\ge\frac{2}{100}.
        \end{cases}
\end{equation}
We obtain a stress--strain curve whose gradient increases each time a new fibril becomes taut. This situation is plotted for one, two and five fibrils in Figure \ref{stress}, using a value for $E$ of 5 GPa, as calculated in \cite{WBHM07} for a rat tail tendon. It can be seen how the curve becomes smoother as more fibrils are added. In reality, ligaments and tendons consist of tens of thousands of fibrils of different lengths, thus giving a smooth curve such as that shown in Figure \ref{typical}.

\section*{Discussion}
\label{sec:discussion}

The physical devices described in this paper are simple to build and operate. Each device illustrates a simple mathematical model of a physical phenomenon in the human body and can be used to kick-start discussions about mathematics and its applications. We have found that secondary school students find the devices engaging and tend to be surprised that mathematics has such `real world' applications that are often not discussed in school. The mathematical models contain a range of concepts with differing degrees of complexity, from simple equations such as Ohm's law and Hooke's law to the differential equations governing fluid flow, which allows them to appeal to a wide age range of students. We hope that this article will inspire readers to create devices of their own in order to aid their outreach efforts and help them to survive science fairs!

\section*{Acknowledgements}
The authors gratefully acknowledge the financial support of the Engineering and Physical Sciences Research Council (EPSRC). P.P. is supported by a Doctoral Prize Fellowship and T.S. is supported by his fellowship grant EP/L017997/1. We would also like to thank Barry Griffiths and Joshua Kliment for help with building the electrical circuits described in the article.

     \begingroup
     \parindent 0pt
     \parskip 2ex
     \def\enotesize{\normalsize}
     \theendnotes
     \endgroup


\begin{thebibliography}{}

\bibitem{aorta}
Belz, G. G. (1995). Elastic properties and Windkessel function of the human aorta. \textit{Cardiovascular Drugs and Therapy}, vol. 9(1), pp. 73--83.

\bibitem{windkesselvid}
\emph{The Windkessel principle visualised,}\\ https://www.youtube.com/watch?v=Bx9Nu2PkPsE

\bibitem{ohms}
\emph{Voltage, Current, Resistance, and Ohm's Law,}\\ https://learn.sparkfun.com/tutorials/voltage-current-resistance-and-ohms-law

\bibitem{simulation}
\emph{Full-wave rectifier with filter,} http://www.falstad.com/circuit/e-fullrectf.html

\bibitem{circuit}
\emph{The Smoothing Capacitor,}\\http://www.electronics-tutorials.ws/diode/diode\textunderscore 6.html

\bibitem{windkessel2}
Catanho, M., Sinha, M., and Vijayan, V. (2012). Model of Aortic Blood Flow Using the Windkessel Effect. \emph{Project Report}, University of California San Diego.

\bibitem{RW15}
Reese, S.P. and Weiss, J.A. (2015). Tendons and ligaments: current state and future directions. In: De, S., Hwang, W. and Kuhl, E. ed. \textit{Multiscale modelling in biomechanics and mechanobiology}. Springer-Verlag London, pp. 159--206.

\bibitem{S15}
Shearer, T. (2015). A new strain energy function for the hyperelastic modelling of ligaments and tendons based on fascicle microstructure, \textit{J. Biomech.}, vol. 48, pp. 290--297.

\bibitem{KPB80}
Kastelic, J., Paller, L. and Baer, E. (1980). A structural model for tendon crimping, \textit{J. Biomech.}, vol. 13, pp. 887--893.

\bibitem{WBHM07}
Wenger, M.P.E., Bozec, L., Horton, M.A. and Mesquida, P. (2007). Mechanical properties of collagen fibrils, \textit{Biophys. J.}, vol. 93, pp. 1255--1263.

\end{thebibliography}
\end{document}